# Optical tweezers with enhanced efficiency based on laser-structured substrates

D. G. Kotsifaki[1], M. Kandyla[2], I. Zergioti[1], M. Makropoulou[1], E. Chatzitheodoridis[3], A. A. Serafetinides[1,a)]

[1]*Physics Department, National Technical University of Athens, Heroon Polytechniou 9, 15780 Athens, Greece*

[2]*Theoretical and Physical Chemistry Institute, National Hellenic Research Foundation, 48 Vasileos Constantinou Avenue, 11635 Athens, Greece*

[3]*School of Mining and Metallurgical Engineering, National Technical University of Athens, 15780 Athens, Greece*

---

[a)] Electronic mail: aseraf@central.ntua.gr

**ABSTRACT**

We present an optical nanotrapping setup that exhibits enhanced efficiency, based on localized plasmonic fields around sharp metallic features. The substrates consist of laser-structured silicon wafers with quasi-ordered microspikes on the surface, coated with a thin silver layer. The resulting optical traps show orders of magnitude enhancement of the trapping force and the effective quality factor.

Conventional optical tweezers, initially proposed by Ashkin[1], are widely employed for trapping micro- and nanometer-size particles, from micro-colloids and droplets[2,3] to living cells[4,5], in a microfluidic environment. This technique is based on radiation pressure forces, and stable trapping is achieved when the gradient of the electromagnetic field intensity (gradient force) dominates scattering forces and Brownian motion[1]. However, in the case of nanometer-size particles, the magnitude of the gradient force decreases. Additionally, damping of the trapped particles decreases rapidly due to a decline in viscous drag[6]. The above mentioned effects work against stable trapping. Therefore, optical trapping of nanometric particles requires more sophisticated optical techniques[7]. Alternative methods based on evanescent fields are used for optical nanotrapping, such as scattering by sub-wavelength particles[8,9] or confinement of light in a metallic tip[10]. Surface plasmons have also been employed for trapping particles in the Rayleigh regime[11,12]. The leading experimental observation of plasmonic trapping involved a glass surface decorated by micrometer-size gold disks[13,14]. Precise manipulation of trapped nanoparticles can be achieved through plasmonic antennas, which have the ability to concentrate and propagate light beyond the diffraction limit[15-17]. In 2008, Grigorenko *et al.* reported measurements of optical forces of about 2 nN on polystyrene particles, using gold nanopillars[18]. Later, Righini *et al.* reported parallel trapping of nanoparticles, using unfocused laser irradiation on top of an array of gap antennas formed by adjacent gold nanobars[19]. Plasmonic traps may play a key role in the development of applications, such as lab-on-a-chip, with increased functionalities. Compared to conventional optical tweezers, plasmonic traps do not require elaborate optical setups to create the trapping volume. Lately, plasmonic traps have become an important tool for immobilizing nanoparticles or cells on a substrate.

In this work, we report the fabrication of an optical trap with enhanced efficiency, based on localized plasmonic fields around sharp metallic features. The trap is created near the surface of a laser-structured silicon substrate coated with a thin silver film. A schematic representation of the optical trap is shown in Fig. 1. Polystyrene beads of 900 nm diameter (Sigma, Aldrich), suspended in deionized water of refractive index n = 1.33, were trapped near the focal point of a continuous wave Nd:YAG laser operating at 1.064 μm in TEM$_{00}$ mode, with maximum output power P = 500 mW. The laser beam was introduced into an optical microscope and focused by a 100x (NA = 1.25) oil-immersion objective lens onto the substrate. The substrates were placed on a motorized x-y translation stage (Standa 8MT 167-2S) and were illuminated by white light so that the image of the trapped beads could be captured on a CCD camera. The beads were trapped and manipulated by the focused laser beam at 1 μm distance from the surface of the substrates. This distance was measured each time by bringing into focus first the peak of the spikes and then the trapped beads and subsequently calculating their distance *via* the relevant microscope software.

Figures 2(a) – 2(b) show scanning electron microscope (SEM) images of the laser-structured substrates. We prepared microstructured samples by irradiating (111)-oriented silicon wafers with nanosecond laser pulse trains, at normal incidence in SF$_6$ atmosphere. The laser source creating the microstructuring was the 4$^{th}$ harmonic of a Q-switched Nd:YAG laser system (wavelength at 266 nm, 4 ns pulse duration). The laser pulses were focused by a 20-cm focal length lens behind the sample surface. The sample was placed inside a vacuum chamber filled with the electronegative gas SF$_6$ at 500 mbar pressure. A train of 1000 pulses irradiated each spot on the silicon wafer in order to microstructure the surface. The morphology of the microstructured spots depends strongly on the incident laser fluence. Figure 2(a) shows a silicon spot irradiated by laser pulses of fluence 0.96 J/cm$^2$, while Fig. 2(b) shows a silicon spot irradiated by lower-fluence laser pulses. The spot in Fig. 2(a) shows well-formed, quasi-ordered microspikes on the surface, while the spot in Fig. 2(b) shows a rippled surface without sharp features. The spikes in Fig. 2(a) are conical, with tip diameter

of 1-3 μm and height ~25 μm. The distance between individual spikes is 8-15 μm. The focal point of the laser beam in the optical trap setup has a diameter of $2\lambda/(\pi\,NA) \approx 540$ nm, therefore it illuminates only one microspike. The microstructured silicon wafers were coated with an 80-nm silver layer using thermal evaporation, in order to be used as substrates for optical trapping. Some wafers were not subjected to silver deposition, in order to serve as reference samples.

In order to determine the effective optical trap force, we performed escape velocity measurements. By translating the microstructured substrates at a constant velocity under the focal point of the trapping laser beam, we were able to measure the velocity (escape velocity) for which the polystyrene beads escaped the optical trap. At the escape velocity, the effective trapping force is considered to be equal to the viscous drag force described by the modified Stokes law,

$$F = K \cdot 6\pi\eta r v_{esc} \qquad (1)$$

where $\eta$ is the water viscosity, $r$ the bead radius, $v_{esc}$ the escape velocity of the bead and $K$ a dimensionless correction coefficient[18]. Figure 3 shows the effective force exerted on 900-nm polystyrene beads, optically trapped above both microstructured and flat substrates, with or without a thin silver layer, as a function of the power of the trapping laser beam. In this graph, we can observe that the trapping force changes significantly with the substrate morphology. The substrate with well-formed silicon spikes (shown in Fig. 2(a)), coated with a thin silver layer, shows two orders of magnitude enhancement of the trapping force, compared to the other samples. We note that the combination of the microspikes and the silver layer is necessary in order to observe enhancement of the trapping force. When we induce trapping above a flat silicon substrate, coated with a thin silver layer, the trapping force is much lower. We obtain similar results when we induce trapping above a silicon substrate with not well-formed spikes (shown in Fig. 2(b)), coated with a thin silver layer.

Additionally, when we induce trapping above a silicon substrate with well-formed spikes but without the thin silver layer, the trapping force remains low. The inset in Fig. 3 shows the effective force for the substrates that didn't show enhanced trapping, plotted on a finer scale. It is interesting to note that the rippled silicon substrate (with not well-formed spikes), coated with a thin silver layer, forms a weaker trap, compared with the other two substrates shown in the inset in Fig. 3. This difference may be explained by destructive interference effects, caused by neighboring silver ripples, which reduce the effective trapping force. The maximum trapping force of 4.3 nN, measured on the silicon substrate with well-formed microspikes coated with a thin silver layer, is about 100 times higher than the force we obtain with a conventional optical trap on glass, indicating the substrates we used achieve orders of magnitude enhancement of the trapping efficiency.

Knowing the trapping force, $F$, we can evaluate the effective quality factor of the optical trap, $Q$. The effective quality factor is defined as

$$Q = Fc/nP \qquad (2)$$

where $c$ is the speed of light, $n$ is the refractive index of the surrounding medium and $P$ is the power of the trapping laser beam. The quality factors for the samples used in this study are given in Table I. We note that these are average values and should be regarded as approximations, since for some of the samples the optical force is a nonlinear function of the trapping laser power, as shown in Fig. 3. In previous studies[20], it was observed that increasing the trapping laser power changed the axial trap location, hence a single value for the quality factor could not be derived for a given particle over all laser powers. The $Q$ values in Table I demonstrate that the microstructured silicon substrate, with well-formed spikes, covered by a thin silver layer, shows a two orders of magnitude enhancement of the quality factor, compared to the other substrates. A possible explanation for the enhanced efficiency of the described optical trap is based on the action of localized plasmonic fields, created around the sharp silicon microspikes covered by silver nanoparticles. Similar behavior was observed with femtosecond laser microstructured and nanostructured silicon substrates coated with a thin silver layer that have been shown to induce enhanced light absorption[21] and surface-enhanced Raman scattering (SERS)[22], respectively. In the case of SERS, the enhancement of the signal was attributed to large local field enhancements, produced by silver nanoparticles that form on the silicon nanospikes during silver deposition[22]. The formation of silver nanoparticles, instead of a smooth silver film, was favored by the nanometric roughness of the surface of the laser-structured spikes. Similarly, light absorption enhancement on silver-coated silicon microspikes was attributed to plasmon excitation induced by the formation of periodic metallic structures and the presence of metallic nanoparticles on the spike surface[21]. Figure 2(a) shows the quasi-periodic microspikes formed in this work, while the inset in Fig. 2(a) shows a magnified view of individual spikes covered by silver nanoparticles. Nanosecond laser fabricated silicon microspikes have also been found to have nanometric roughness on their surface[23-25], favoring the formation of nanoparticles during silver deposition. The combination of quasi periodic metallic structures with metallic nanoparticles results in electromagnetic field enhancement due to plasmon resonances during optical trapping and increases the quality factor of the trap by two orders of magnitude. The remarkably high quality factor of $Q = 9.86 \pm 0.12$, obtained with silver-coated silicon microspikes in this work, is comparable to quality factors obtained with other plasmonic optical traps[18].

In conclusion, we have demonstrated an optical tweezer system based on a simple process of fabricating microstructured silicon substrates with a solid state nanosecond UV laser, coated with a thin silver layer. The substrate morphology creates a strong plasmonic optical trap and offers a remarkably large trapping efficiency. This optical tweezers system offers additional potential for plasmonic trapping and paves the way for applications in nanoengineering and nanobiology.

**FIGURE CAPTIONS**

**Figure 1.** Schematic diagram of the optical trap based on microstructured silicon substrates.

**Figure 2.** (a),(b) SEM images (viewed at $45^o$) of laser microstructured silicon substrates, corresponding to different laser fluences. Inset: Magnified view of a silver-coated substrate.

**Figure 3.** Trapping force on 900-nm polystyrene beads as a function of laser power for various substrates. The different substrates are: a) the laser microstructured silicon sample shown in Fig. 2a with well-formed microspikes, coated with a 80-nm silver layer (squares), b) the laser microstructured silicon sample shown in Fig. 2b, coated with a 80-nm silver layer (circles), c) a flat, unprocessed silicon wafer, coated with a 80-nm silver layer (triangles), and d) a laser microstructured silicon sample with well-formed microspikes, without a silver layer (diamonds).

**Table I**. Effective trapping quality factor, *Q*, for various substrates.

**TABLE I**

| Substrate | Quality Factor *Q* |
|---|---|
| Well-formed silicon microspikes, coated with silver layer | 9.860 ± 0.120 |
| Rippled silicon surface, coated with silver layer | 0.008 ± 0.000 |
| Flat silicon wafer, coated with silver layer | 0.043 ± 0.005 |
| Well-formed silicon microspikes, without silver layer | 0.059 ± 0.003 |
| Glass | 0.018 ± 0.001 |

FIGURES

Figure 1

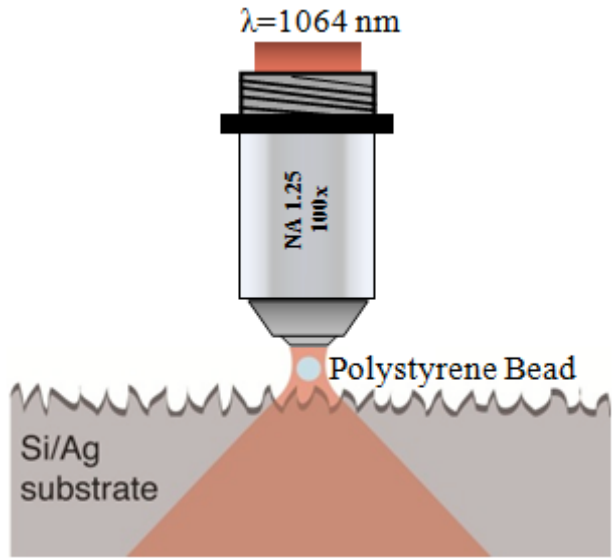

Figure 2 (a) - (b)

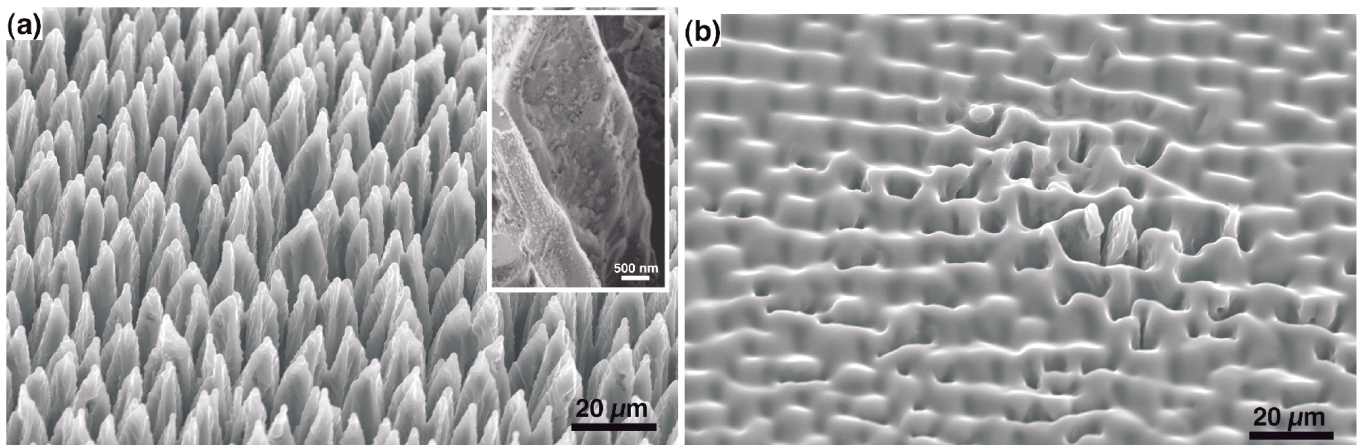

Figure 3